\documentclass{article}
\usepackage{amsmath}
\usepackage{amssymb}
\usepackage{color}
\usepackage{graphicx}
\begin{document}
\title{Time and Evolution in Quantum and Classical Cosmology}

\author{Alexander Yu. Kamenshchik $^{1}$, Jeinny Nallely P\'erez Rodr\'iguez$^{2}$\\
 and Tereza Vardanyan $^{3}$}
 \date{}
\maketitle

$^{1}$  Dipartimento di Fisica e Astronomia dell'Universit\`a  di Bologna and INFN, Via Irnerio 46, 40126 Bologna, Italy and L.D. Landau Institute for Theoretical Physics of the Russian Academy of Sciences, Kosygin Street. 2, 119334 Moscow, Russia; kamenshchik@bo.infn.it\\
$^{2}$  Dipartimento di Fisica e Astronomia dell'Universit\`a  di Bologna, Via Irnerio 46, 40126 Bologna, Italy; jeinnynallely.perez@studio.unibo.it\\
$^{3}$  Dipartimento di Fisica e Chimica, Universit\`a di L'Aquila, 67100 Coppito, L'Aquila, Italy and INFN, Laboratori Nazionali del Gran Sasso, 67010 Assergi, L'Aquila, Italy; tereza.vardanyan@aquila.infn.it\\

\abstract{We analyze the issue of dynamical evolution and time in quantum cosmology. We emphasize the problem of choice of phase space variables that can play the role of a time parameter in such a way that for expectation values of quantum operators the classical evolution is reproduced. We show that it is neither necessary nor sufficient for the Poisson bracket between the time variable and the super-Hamiltonian to be equal to unity in all of the phase space. We also discuss the question of switching between different internal times as well as the Montevideo interpretation of quantum theory.}


\section{Introduction}

The problem of ``disappearance of time'' in quantum gravity and cosmology is well known and has a rather long history (see, e.g., Refs. \cite{Kuchar,Kiefer,Ryan} and Refs therein). Let us briefly recall the essence of the problem. Not all of the ten Einstein equations in the four-dimensional spacetime contain the second time derivatives: four of them include only the first time derivatives and are treated as constraints. The presence of such constraints is connected with the fact that the action is invariant under the spacetime diffeomorphisms group. To~treat this situation, it is rather convenient to use the Arnowitt-Deser-Misner formalism~\cite{ADM}. 
 The Hilbert-Einstein Lagrangian of General Relativity contains the Lagrange multipliers $N$ and $N^i$, which are called lapse and shift functions. There are also the dynamical degrees of freedom connected with the spatial components of the metric $g_{ij}$ and with the non-gravitational fields present in the universe. In order to use the canonical formalism, one introduces the conjugate momenta and makes a Legendre transformation. Then, one discovers that the Hamiltonian is proportional to the linear combination of constraints multiplied by the Lagrange multipliers~\cite{DeWitt}. Thus, the Hamiltonian vanishes if the constraints are satisfied. This can be interpreted as an impossibility of writing down a time-dependent Schr\"odinger equation. This problem can be seen from a somewhat different point of view. If one applies 
the Dirac quantization procedure~\cite{Dirac}, then the constraint in which the classical phase variables are substituted by the quantum operators should annihilate the quantum state of the system under consideration. Gravitational constraints contain momenta, which classically are time derivatives of fields, but their connection to the classical notion of time vanishes in the quantum theory, where momenta are simply operators satisfying commutation relations. The main constraint arising in General Relativity 
is quadratic in momenta and gives rise to the Wheeler-DeWitt equation~\cite{DeWitt, Wheeler}. It is time-independent, but we know that the universe lives in time. Where is it hidden? 
 
Generally, all the explanations of the reappearance of time are based on the identification of some combination of the phase space variables of the system under consideration (in our case, the whole universe) with the time parameter. Then, the Wheeler-DeWitt equation is transformed into the Schr\"odinger-type equation with respect to this time parameter. As is well known, in gauge theories or, in other words, in theories with first class constraints, one should choose the gauge-fixing conditions (see, e.g., Ref. \cite{Sundermeyer, Henneaux}). If we choose a time-dependent classical gauge-fixing condition, we can define a classical time parameter expressed by means of a certain combination of phase variables. Some other variables are excluded from the game by resolving the constraints. As a result, one obtains a non-vanishing effective physical Hamiltonian that depends on the remaining quantized phase space variables. This Hamiltonian governs the evolution of the wave function describing the degrees of freedom not included in the definition of time. Then, one can reconstruct a particular solution of the initial Wheeler-DeWitt equation, corresponding to the solution of the physical Schr\"odinger equation. The above procedure was elaborated in detail in Ref. \cite{Barv}, and, in Ref. \cite{Bar-Kam}, it was explicitly applied to some relatively simple cosmological models. It is important to note that, in this approach, the time parameter and the Hamiltonian are introduced before the quantization of the physical degrees of freedom. 

Another approach is based on the Born-Oppenheimer method~\cite{Brout, Brout-Venturi,Venturi,we-BO}, which was very effective in the treatment of atoms and molecules~\cite{BO}. 
This method relies on the existence of two time or energy scales. In the case of cosmology, one of these scales is related to the Planck mass and to the slow evolution of the gravitational degrees of freedom. The other one is characterized by a much smaller energy and by the fast motion of the matter degrees of freedom. 
The solution to the Wheeler-DeWitt equation is factorized into a product of two wave functions. One of them has a semiclassical structure with the action satisfying the Einstein-Hamilton-Jacobi equation and the expectation value of the energy-momentum tensor serving as the source for gravity; the semiclassical time arises from this equation. The second wave function describes quantum degrees of freedom of matter and satisfies the effective Schr\"odinger equation. However, the principles of separation between these two kinds of degrees of freedom in the two approaches described above are somewhat different. The comparison of these approaches and attempts to establish correspondence between them were undertaken in Ref. \cite{we-IJMPD,Chat,Chat1,Chat2}. 

We have already said that the very concept of time is in some sense classical. However, the time parameter, which arises when we work with the Wheeler-DeWitt equation, is a function of the phase space variables, which are quantum operators. Thus, we should ``freeze'' the quantum nature of these variables when using them to introduce time, and the time parameter should be such that expectation values of operators evolve classically. Here, we arrive to the main question of this review, which was stimulated by the relatively unknown paper by Asher Peres~\cite{Peres} with a rather significant title ``Critique of the Wheeler-DeWitt equation''. While we do not agree with the conclusions of this paper, we find it thought-provoking and worth analyzing. The author of Ref. \cite{Peres} assumes that the time parameter should be chosen in such a way that the classical equations of motion are satisfied not only on the constraint surface but in all of the phase space. We will show that this condition is neither necessary nor sufficient. 

In Ref. \cite{Peres}, just like in our papers~\cite{Bar-Kam, we-IJMPD}, a very simple model was considered---a flat Friedmann universe filled with a minimally coupled scalar field. Despite its simplicity, this model appears to be rather instructive. In the recent paper, Ref. \cite{Hohn}, it is studied in detail in the framework of the relational approach to quantum theory and quantum gravity developed in the series of preceding papers~\cite{Hohn1,Hohn2,Hohn3,Hohn4}. In Ref. \cite{Hohn5,Hohn6}, the ideas of the relational or quantum reference  frame approach are further developed. These papers attracted attention of the authors of Ref.\cite{Mont}, where their connection with the Montevideo interpretation of quantum theory is studied~\cite{Mont0,Mont1,Mont2,Mont3,Mont4}.

The structure of this paper is the following: In the next section, we present for completeness some basic formulae of the Arnowitt-Deser-Misner formalism and the Dirac quantization of gravity, arriving to the general form of the Wheeler-DeWitt equation. In the third section, we present a simple toy model, which was analyzed by us in Refs. \cite{Bar-Kam,we-IJMPD}, as well as by Peres in Ref. \cite{Peres}. Section~\ref{3.1} contains the description of the classical dynamics of the model---a flat Friedmann universe filled with a minimally coupled massless scalar field. In Section~\ref{3.2}, we present the Hamiltonian formalism, Wheeler-DeWitt equation and the gauge fixing procedure for this model in the spirit of Ref. \cite{Barv,Bar-Kam, we-IJMPD}. As is done in Ref. \cite{Peres}, in~Section~\ref{3.3}, we introduce the time parameter by using direct analysis of the Poisson brackets, without the super-Hamiltonian constraint. We show that, in Ref. \cite{Peres}, an essential additional free parameter was overlooked and that our approach in Ref. \cite{Barv,Bar-Kam,we-IJMPD} corresponds to one choice of this parameter, whereas, in Ref. \cite{Peres}, a different choice was made. In Section~\ref{3.4}, we reproduce the results of 3.3 by using a canonical transformation and constructing the corresponding generating function. In Section~\ref{3.5}, we show that only our choice of the free parameter and the explicit implementation of the super-Hamiltonian constraint permits to obtain the quantum evolution that is in agreement with the classical one. Section 4 contains the comparison of our results with those obtained in Ref. \cite{Hohn} and the discussion of some related questions. In the last section, we discuss the obtained results and possible future directions of research.

\section{The Arnowitt-Deser-Misner Formalism, Dirac Quantization and the Wheeler-DeWitt Equation}

To make the transition from the Lagrangian formalism to the Hamiltonian formalism, one should break the relativistic covariance and introduce a variable that will play the role of time. The 3+1 foliation of the pseudo-Riemannian manifold serves exactly this purpose (see, e.g., Ref. \cite{3-plus-1} for a very clear and detailed exposition of the 3+1 formalism). 
Constructing 3+1 foliation of the spacetime begins with an embedding of a three-dimensional hypersurface into a four-dimensional spacetime. Such an embedding can be realized by defining a scalar function $t(p)$ on a four-dimensional manifold
(here, $p$ denotes a point on the manifold). A hypersurface can then be defined as a level surface of this function. The gradient of $t(p)$ is a one-form that annihilates the vectors tangent to this hypersurface. The vector field obtained from the contraction of this gradient with the contravariant spacetime metric (we can call it $\vec{\nabla} t$) is orthogonal to the tangent spaces of the hypersurface. If this hypersurface is spacelike, then the orthogonal vector is timelike and can be normalized. The normalized vector field is called the unit normal vector field and is denoted by $\vec{n}$.

We can now define the three-dimensional induced metric $\gamma_{ij}$ on the hypersurface by using the pullback operation of the spacetime metric $g_{\mu\nu}$ to the three-dimensional hypersurface. The induced metric will play the role of phase space variables in the Hamiltonian formalism of General Relativity. On the hypersurface, we define a covariant derivative $D$ associated with the induced three-metric $\gamma_{ij}$. Then, one can introduce the corresponding three-dimensional Riemann--Christoffel curvature tensor, which describes the internal geometry of the hypersurface. To describe the position of a three-dimensional hypersurface inside a four-dimensional spacetime, we also define an extrinsic curvature tensor. To this end, one introduces the Weingarten transformation that maps a tangent vector $\vec{u}$ into the covariant derivative $\nabla_{\vec{u}}\vec{n}$ (associated with the metric $g_{\mu\nu}$) of $\vec{n}$ along $\vec{u}$. To show that the vector $\nabla_{\vec{u}}\vec{n}$ is also tangent to the hypersurface, it is enough to prove that it is orthogonal to the normal vector $\vec{n}$. In this section, it is convenient for us to use the spacetime signature $(-,+,+,+)$. Then, the unit normal vector satisfies 
the relation $\vec{n}^2 = -1$; hence,
$$
\vec{n}\cdot \nabla_{\vec{u}}\vec{n} =\frac12 \nabla_{\vec{u}}(\vec{n}^2) = 0.
$$
 contracting the Weingarten operator with the induced metric, one obtains a symmetric tensor, which we shall call the extrinsic curvature $K_{ij}$:
\begin{equation}
K_{ij}v^iu^j \equiv -\gamma_{ij}v^i(\nabla_{\vec{u}}\vec{n})^{j},
\label{extrinsic}
\end{equation} 
where $\vec{v}$ and $\vec{u}$ are tangent vectors, and $i$, $j$ run over the spatial coordinates. In what follows, Greek indices run over all spacetime coordinates, whereas Latin indices run only over the spatial coordinates. As will be shown, the extrinsic curvature is related to the time derivatives of the induced metric and the corresponding conjugate momenta and, hence, plays a very important role in the Hamiltonian formalism.
 
The next step is the construction of the projector from the spacetime to the hypersurface. This projector transforms any vector $\vec{u}$ into the vector $\vec{u}+(\vec{u}\cdot\vec{n})\vec{n}$. 
In the index notation, this projector can be represented as 
\begin{equation}
\gamma^{\alpha}_{\beta} = \delta^{\alpha}_{\beta} + n^{\alpha}n_{\beta}.
\label{proj}
\end{equation}
 can easily see that \eqref{proj} acts as an identity operator for vectors tangent to the hypersurface, whereas it annihilates vectors orthogonal to the hypersurface. Obviously, the projector can act not only on vectors, but also on arbitrary tensors. 
We can also introduce an extension of tensors defined on the hypersurface by treating them as a tensors defined in the spacetime. The principle is the following: the contraction of the extended tensor with vectors should give the same result as the contraction of the original one if the vectors are tangent to the hypersurface and give zero if at least one of the vectors is orthogonal to the hypersurface. 

Interestingly, the extended version of the induced metric can be obtained from \eqref{proj} by lowering its index with the spacetime metric,
\begin{equation}
\gamma_{\alpha\beta} = g_{\alpha\beta}+n_{\alpha}n_{\beta}.
\label{ind}
\end{equation}
 the extrinsic curvature, the extended expression has the following form:
\begin{equation}
K_{\alpha\beta} = -\nabla_{\beta}n_{\alpha}-a_{\alpha}n_{\beta},
\label{ext1}
\end{equation}
where the acceleration vector $\vec{a}$ is defined as $\vec{a} \equiv \nabla_{\vec{n}}\vec{n}$.

Using the projector \eqref{proj}, one can find the difference between the covariant derivative $\nabla$ and 
the covariant derivative $D$ acting on the vector fields tangent to the hypersurface:
\begin{equation}
D_{\vec{u}}\vec{v} = \nabla_{\vec{u}}\vec{v}+K_{ij}v^iu^j\vec{n}.
\label{connection}
\end{equation}
 difference is directed along the vector normal to the hypersurface and depends on the extrinsic curvature.

Now, we can try to express the four-dimensional curvature tensor in terms of the three-dimensional Ricci tensor and the extrinsic curvature. The corresponding equations are called Gauss-Codazzi relations. 
Using the Ricci identity for the covariant derivative $D$,
\begin{equation}
D_{\mu}D_{\nu}u^{\alpha}-D_{\nu}D_{\mu}u^{\alpha} ={}^{(3)} R^{\alpha}_{\beta\mu\nu}u^{\beta},
\label{Ricci}
\end{equation}
the analogous identity for the covariant derivative $\nabla$ and the projector \eqref{proj}, one can show that 
\begin{equation}
\gamma_{\alpha}^{\mu}\gamma_{\beta}^{\nu}\gamma_{\rho}^{\gamma}\gamma_{\delta}^{\sigma}R^{\rho}_{\sigma\mu\nu}=
{}^{(3)}R^{\gamma}_{\delta\alpha\beta}+K^{\gamma}_{\alpha}K_{\delta\beta}-K^{\gamma}_{\beta}K_{\alpha\delta},
\label{Gauss}
\end{equation}
where ${}^{(3)}R^{\gamma}_{\delta\alpha\beta}$ is the Riemann curvature tensor (also called intrinsic curvature) associated with the induced metric $\gamma_{\alpha\beta}$. By contracting this identity with respect to the indices $\alpha$ and $\gamma$, we obtain the following relation:
\begin{equation}
\gamma^{\mu}_{\alpha}\gamma_{\beta}^{\nu}R_{\mu\nu}+\gamma_{\alpha\mu}\gamma_{\beta}^{\rho}n^{\nu}n^{\sigma}R^{\mu}_{\nu\rho\sigma}={}^{(3)}R_{\alpha\beta}+KK_{\alpha\beta}-K_{\alpha\mu}K_{\beta}^{\mu},
\label{Gauss1}
\end{equation}
where $K \equiv  K_{\mu}^{\mu}$. Now, we can contract this relation with the contravariant induced metric $\gamma^{\alpha\beta}$ and obtain
\begin{equation}
R+2n^{\mu}n^{\nu}R_{\mu\nu}={}^{(3)}R+K^2-K_{i}^{j}K_{j}^{i},
\label{Gauss2}
\end{equation}
where we used that $K_{\mu}^{\nu}K_{\nu}^{\mu}=K_{i}^{j}K_{j}^{i}$.

We can also apply the Ricci identity to the vector $n^{\alpha}$ and project the result on the hypersurface; this results in
\begin{equation}
\gamma_{\rho}^{\gamma}\gamma_{\alpha}^{\mu}\gamma_{\beta}^{\nu}n^{\sigma}R^{\rho}_{\sigma\mu\nu}=D_{\beta}K^{\gamma}_{\alpha}-D_{\alpha}K^{\gamma}_{\beta}.
\label{Codazzi}
\end{equation} 
 contracting this relation with respect to the indices $\alpha$ and $\gamma$, we obtain 
\begin{equation}
\gamma_{\beta}^{\nu}n^{\sigma}R_{\sigma\nu}=D_{\beta}K-D_{\nu}K_{\beta}^{\nu}.
\label{Codazzi1}
\end{equation}
 
An embedding of a three-dimensional spacelike hypersurface into a four-dimensional spacetime is not sufficient for the construction of the Hamiltonian formalism: one should also know the time evolution of the embedded hypersurface. In other words, we need the complete 3+1 foliation of the spacetime. Thus, we need to introduce a family of spacelike hypersurfaces that cover the whole spacetime and are parametrized by the parameter $t$, which we shall later identify with time. 
We start by defining the famous lapse function $N$ (this name was invented by J.A. Wheeler in 1964~\cite{Wheeler-lapse}):
\begin{equation}
\vec{n} = -N\vec{\nabla} t.
\label{lapse0}
\end{equation}
The lapse function plays a very important role in both the 3+1 foliation of a four-dimensional spacetime and in the Hamiltonian formalism of General Relativity. It is also convenient to introduce another vector field orthogonal to the spacelike hypersurfaces of the foliation, the normal evolution vector $\vec{m}$:
\begin{equation}
\vec{m} = N\vec{n}.
\label{m}
\end{equation}
If we start at the hypersurface parameterized by $t$ and move along an integral curve of the vector field $\vec{m}$, denoting the change of the curve's parameter by $\delta t$, we will find ourselves at the hypersurface parameterized by $t+\delta t$. One can obtain different useful relations involving the vector fields $\vec{n}$ and $\vec{m}$ and the lapse function 
$N$. The acceleration $a$ can be expressed as
\begin{equation}
a_{\alpha} = D_{\alpha}\ln N.
\label{acc-lapse}
\end{equation}
The Lie derivative of the induced metric along the normal evolution field is 
\begin{equation}
L_{\vec{m}}\gamma_{\alpha\beta}=-2NK_{\alpha\beta},
\label{Lie-m}
\end{equation}
while the Lie derivative of the induced metric along the unit normal field is 
\begin{equation}
L_{\vec{n}}\gamma_{\alpha\beta}=-2K_{\alpha\beta}.
\label{Lie-n}
\end{equation}

Now, we can complete the expression for $R$ in terms of ${}^{(3)}R$, the extrinsic curvature and the lapse function. We apply the Ricci identity to the normal unit vector and project the result two times on the hypersurface and one time on the normal direction. As a result, we obtain
\begin{equation}
\gamma_{\alpha\mu}\gamma_{\beta}^{\nu}n^{\rho}n^{\sigma}R^{\mu}_{\rho\nu\sigma}=\frac{1}{N}L_{\vec{m}}K_{\alpha\beta}+\frac{1}{N}D_{\alpha}D_{\beta}N+K_{\alpha\mu}K_{\beta}^{\mu}.
\label{Gauss3}
\end{equation} 
Combining this equation with Equation \eqref{Gauss1}, we obtain
\begin{equation}
\gamma_{\alpha}^{\mu}\gamma_{\beta}^{\nu}R_{\mu\nu}=-\frac{1}{N}L_{\vec{m}}K_{\alpha\beta}-\frac{1}{N}D_{\alpha}D_{\beta}N
+{}^{(3)}R_{\alpha\beta}+KK_{\alpha\beta}-2K_{\alpha\mu}K_{\beta}^{\mu}.
\label{Gauss4}
\end{equation}
Taking the trace of this equation with respect to the metric $\gamma$ gives
\begin{equation}
R+n^{\mu}n^{\nu}R_{\mu\nu}={}^{(3)}R+K^2-\frac{1}{N}L_{\vec{m}}K-\frac{1}{N}D_{i}D^{i}N.
\label{Gauss5}
\end{equation}
Now, we can combine this equation with Equation \eqref{Gauss2} and arrive at the following expression:
\begin{equation}
R={}^{(3)}R+K^2+K_{ik}K^{ij}-\frac{2}{N}L_{\vec{m}}K-\frac{2}{N}D_{i}D^{i}N.
\label{Gauss6}
\end{equation}

However, to transition to the Hamiltonian formalism, one should use the explicit time derivatives of field variables. Thus, we shall identify the parameter $t$, defining the hypersurfaces of the 3+1 foliation, with time. Then, we can introduce the vector field $\frac{\partial}{\partial t}$. This vector field is tangent to the integral curves, which are the curves along which 
the three spatial coordinates are constant. The action of this vector field on the gradient of the function defining 3+1 foliation is equal to unity, but, at the same time, it is not necessarily orthogonal to the spacelike hypersurfaces. Thus, we can write the following identity
\begin{equation}
\frac{\partial}{\partial t} = N\vec{n} + \vec{N},
\label{shift}
\end{equation}
where $\vec{N}$ is the shift vector field. This vector field is orthogonal to the normal and the components of the metric in terms of the coordinates $t$ and $x_{i}$, where $i=1,2,3$, are given by the expression
\begin{equation}
ds^2=-(N^2-N_iN^i)dt^2+2N_idtdx^i+\gamma_{ij}dx^idx^j.
\label{ADM}
\end{equation}
The components of the contravariant metric are 
\begin{equation}
g^{00} = -\frac{1}{N^2},\ g^{i0}=\frac{N^i}{N^2},\ g^{ij} = \gamma^{ij}-\frac{N^iN^j}{N^2}.
\label{ADM1}
\end{equation}
The determinant of the four-dimensional metric is 
\begin{equation}
g = -N^2\gamma.
\label{deter}
\end{equation}

From Equation \eqref{shift}, it follows that the Lie derivative of a tensor field along the normal evolution vector $\vec{m}$ is 
\begin{equation}
L_{\vec{m}}T = L_{\frac{\partial}{\partial t}}T-L_{\vec{N}}T.
\label{shift1}
\end{equation}
Using \eqref{shift1}, it is easy to show that 
\begin{equation}
L_{\vec{m}}K_{ij} = \left(\frac{\partial}{\partial t}-L_{\vec{N}}\right)K_{ij},
\label{shift2}
\end{equation}
and 
\begin{equation}
K_{ij} = -\frac{1}{2N}\left(\frac{\partial \gamma_{ij}}{\partial t} - D_{i}N_j-D_{j}N_i\right).
\label{shift3}
\end{equation}
From Equation \eqref{shift3}, we see that the time derivative of the spatial metric is related to the extrinsic curvature.
Now, the Hilbert-Einstein action 
\begin{equation}
\int d^4x \sqrt{-g}R
\label{H-E}
\end{equation}
can be expressed as 
\begin{equation}
S = \int dt \int d^3x\sqrt{\gamma}N({}^{(3)}R+K_{ij}K^{ij}-K^2).
\label{H-E1}
\end{equation}
We can finally make the Legendre transformation to transition to the Hamiltonian formalism. 
The time derivatives of the lapse and shift functions are not present in the action; thus, they will play the role of Lagrange multipliers. The corresponding conjugate momenta are equal to zero and should remain equal to zero during the evolution. 
The momenta conjugate to the components of the three-metric $\gamma_{ij}$ are defined as usual, 
\begin{equation}
\pi^{ij}=\frac{\partial {\cal L}}{\partial \gamma_{ij}}=\sqrt{\gamma}(K\gamma^{ij}-K^{ij}).
\label{conjug}
\end{equation}
We can resolve Equation \eqref{conjug} with respect to the extrinsic curvature:
\begin{equation}
K_{ij} = \frac{1}{\sqrt{\gamma}}\left(\frac12\gamma_{ij}\gamma_{kl}\pi^{kl}-\gamma_{ik}\gamma_{jl}\pi^{kl}\right).
\label{conjug1}
\end{equation} 
Using \eqref{shift3} and \eqref{conjug1}, we find 
\begin{equation}
\frac{\partial \gamma_{ij}}{\partial t}=D_{i}N_j+D_jN_i+
\frac{N}{\sqrt{\gamma}}(2\gamma_{ik}\gamma_{jl}\pi^{kl}-\gamma_{kl}\gamma_{ij}\pi^{kl}).
\label{conjug2}
\end{equation}
Now, using \eqref{H-E1}--\eqref{conjug2}, we can calculate the Hamiltonian density 
\begin{eqnarray}
&&{\cal H} = \pi^{ij}\frac{\partial \gamma_{ij}}{\partial t} - {\cal L} = NH_{\perp}+N^iH_i, 
\label{Hamilton0}
\end{eqnarray}
where 
\begin{equation}
H_{\perp} = \frac{1}{\sqrt{\gamma}}\left(\gamma{ik}\gamma_{jl}-\frac12\gamma_{ij}\gamma_{kl}\right)\pi^{ij}\pi^{kl}-\sqrt{\gamma}{}^{(3)}R,
\label{super}
\end{equation}
\begin{equation}
H_{i}=-\gamma_{ik}D_{j}\pi^{jk}-\gamma_{ij}D_k\pi^{jk}.
\label{super1}
\end{equation}

The expression \eqref{super} represents the super-Hamiltonian constraint, which has a rather complicated structure.
The expressions \eqref{super1} are the supermomenta constraints and are responsible for the independence of the geometry of the three-dimensional spatial sections of the spacetime on the particular choice of spatial coordinates.
These constraints constitute a system of first class constrains, i.e., the Poisson brackets between them are again proportional to the constraints. Namely, they have the following form:
\begin{eqnarray}
&&\{H_{\perp}(\vec{x}),H_{\perp}(\vec{y})\}=\frac{\partial\delta(\vec{x},\vec{x})}{\partial x^i}(\gamma^{ij}(\vec{x})H_j(\vec{x})+\gamma^{ij}(\vec{y})H_{j}(\vec{y})),\nonumber \\
&&\{H_i(\vec{x}),H_{\perp}(\vec{y})\}=H_{\perp}(\vec{y})\frac{\partial\delta(\vec{x},\vec{y})}{\partial x^i},\nonumber \\
&&\{H_i(\vec{x}),H_j(\vec{y})\}=H_i(\vec{x})\frac{\partial\delta(\vec{x},\vec{y})}{\partial x^i}+H_{i}(\vec{y})\frac{\partial\delta(\vec{x},\vec{y})}{\partial x^j}.
\label{Poisson}
\end{eqnarray}
Remarkably, if we add an action of matter fields to the action \eqref{H-E}, the structure of the Hamiltonian density will not changed and will still be proportional to the linear combination of constraints. The super-Hamiltonian constraint will acquire an additional term equal to the energy density of the corresponding field, while the supermomenta will acquire terms equal to the mixed temporal-spatial components of the corresponding energy-momentum tensor~\cite{Kuchar1,Kuchar2,Kuchar3}.

To quantize the theory, we should substitute all the phase space variables with the corresponding operators and 
the Poisson brackets \eqref{Poisson} with the corresponding commutators. Naturally, such a substitution raises the question of consistency of the quantum version of the relations \eqref{Poisson}. In string and superstring theories, 
analyzing this question resulted in the discovery of the dimensionalities of spacetimes where these theories can be consistently realized. The corresponding analysis in General Relativity is more complicated and the number of works concerning this topic is rather limited. Here, we can mention Ref. \cite{Kam-Lyakh}, where the Hamiltonian BFV-BRST method of the quantization of the constrained systems~\cite{BFV,BFV1,BFV2} was applied to analyze the system \eqref{Poisson} in relatively simple models in quantum cosmology. 

If one considers simple models where it is possible to choose a coordinate system in such a way that the shift functions and the corresponding supermomenta constraints are absent, then the problem of the consistency of the quantized relations \eqref{Poisson} vanishes, and it is necessary to consider only the super-Hamiltonian constraint. As we have already explained in the Introduction, applying the operator version of this constraint to the wave function of the universe gives the Wheeler-DeWitt equation. However, even in this case, we stumble upon some technical and conceptual problems. Thus, the rest of this paper will be devoted to the analysis of these problems based on the consideration of a very simple model---a flat Friedmann universe filled with a massless scalar field.

\section{Flat Friedmann Universe with a Massless Scalar Field\label{3}}

\subsection{Classical Dynamics\label{3.1}}

Let us consider the simplest toy model---a flat Friedmann universe with the metric 
\begin{equation}
ds^2=N^2(t)dt^2 - a^2(t)dl^2,
\label{Fried}
\end{equation} 
where $N$ is, as usual, the lapse function, and $a(t)$ is the scale factor. (In this section, we shall use a different convention for the signature of the spacetime to simplify the comparison with the papers~\cite{we-IJMPD, Peres}.)
This universe is filled with a massless spatially homogeneous scalar field $\phi(t)$ minimally coupled to gravity. 
For this minisuperspace model, the Lagrangian can be written as 
\begin{equation}
 {\cal L} = -\frac{L^3M^2\dot{a}^2a}{2N}+\frac{L^3\dot{\phi}^2 a^3}{2N},
 \label{Lagrange}
 \end{equation}
 where $M$ is a conveniently rescaled Planck mass, and $L$ is a length scale, {introduced to provide the correct dimensionality for the Lagrangian.} 
 It will be convenient to use another parametrization of the scale factor 
 \begin{equation}
 a(t) = e^{\alpha(t)}.
 \label{scale}
 \end{equation}
 Then, 
 \begin{equation}
 {\cal L} = -\frac{L^3M^2\dot{\alpha}^2e^{3\alpha}}{2N}+\frac{L^3\dot{\phi}^2 e^{3\alpha}}{2N}.
 \label{Lagrange1}
 \end{equation} 
 The variation of the Lagrangian (\ref{Lagrange1}) with respect to the lapse function $N$ gives the first Friedmann equation 
 \begin{equation}
 M^2\dot{\alpha^2} = \dot{\phi}^2,
 \label{Fried1}
 \end{equation}
 while its variation with respect to $\phi$ gives the first integral of the Klein-Gordon equation 
 \begin{equation}
 \frac{L^3\dot{\phi}e^{3\alpha}}{N} = p_{\phi} = const.
 \label{KG}
 \end{equation}
 Here, $p_{\phi}$ is the conjugate momentum, which is conserved during the classical time evolution of our universe.
 It will be convenient to choose the cosmic time $t$ as a time parameter, which is equivalent to fixing $N = 1$.
Upon substituting Equation~(\ref{KG}) into Equation~(\ref{Fried1}), we find that, for the expanding universe and $0 \leq t \leq \infty$, 
 \begin{equation}
 e^{3\alpha} = 3\frac{|p_{\phi}|}{ML^3}t,
 \label{expand}
 \end{equation}
 and for the contracting universe, when $-\infty < t \leq 0$,
 \begin{equation}
 e^{3\alpha} = -3\frac{|p_{\phi}|}{ML^3}t.
 \label{contract}
 \end{equation}
 In what follows, we will consider the expanding universe and choose the positive sign for $p_{\phi}$ without losing generality.

\subsection{Hamiltonian formalism, Wheeler-DeWitt Equation, and the Gauge Fixing Procedure\label{3.2}} 

On introducing the conjugate momenta $p_{\phi}$ (see Equation~(\ref{KG})) and 
 \begin{equation}
 p_{\alpha} = -\frac{L^3M^2\dot{\alpha}e^{3\alpha}}{N},
 \label{momentum}
 \end{equation}
 and making the Legendre transformation, we see that the Hamiltonian is 
 \begin{equation}
 {\cal H} = N\left(-\frac{p_{\alpha}^2e^{-3\alpha}}{2M^2L^3}+\frac{p_{\phi}^2e^{-
 3\alpha}}{2L^3}\right) = NH,
 \label{Hamilton10}
 \end{equation}
 where $H$ is nothing but the super-Hamiltonian constraint. From Equations (\ref{Fried1}), (\ref{KG}), and (\ref{momentum}), it is obvious that the Hamiltonian is constrained to vanish:
 \begin{equation}
 H=0.
 \label{constr}
 \end{equation} 
 The action in the Hamiltonian form is 
 \begin{equation}
 S = \int dt (p_{\alpha}\dot{\alpha}+p_{\phi}\dot{\phi}-NH).
 \label{action-Ham}
 \end{equation} 
 
 On performing the procedure of the Dirac quantization of the system with constraints~\cite{Dirac}, we obtain the Wheeler-DeWitt equation:
 \begin{equation}
 \hat{H}|\Psi\rangle = 0. 
 \label{WdW}
 \end{equation}
 Here, the operator $\hat{H}$ arises when we substitute the phase space variables by the corresponding operators and fix some particular operator ordering, and $|\Psi\rangle$ is the quantum state of the universe.
 Now, we shall choose the simplest operator ordering, such that 
 \begin{equation}
 \hat{H} = e^{-3\hat{\alpha}}\left(-\frac{\hat{p}_{\alpha}^2}{2M^2L^3}+\frac{\hat{p}_{\phi}^2}{2L^3}\right).
 \label{WdW1}
 \end{equation}
 It will be convenient to consider the quantum state $|\Psi\rangle$ in the $(\alpha,p_{\phi})$ representation. Thus,
 the Wheeler-DeWitt equation will have the following form:
 \begin{equation}
 \left( \frac{\partial^2}{\partial\alpha^2}+M^2p_{\phi}^2\right)\Psi(\alpha,p_{\phi})=0.
 \label{WdW2}
 \end{equation}
 The general solution of this equation is 
 \begin{equation}
 \Psi(\alpha,p_{\phi}) = \psi_1(p_{\phi_1})e^{iM|p_{\phi}|\alpha}+\psi_1(p_{\phi_2})e^{-iM|p_{\phi}|\alpha}.
 \label{WdW-sol}
 \end{equation}
 
 We are going to derive the effective Schr\"odiger equation for the physical wave function and the physical Hamiltonian 
 following the recipe described in detail in Ref. \cite{Barv,Bar-Kam}. First of all, we have to introduce a time-dependent gauge-fixing condition.
 Let us try to use the following one:
 \begin{equation}
 \xi(\alpha,p_{\alpha},t) = \frac{L^3M^2e^{3\alpha}}{3p_{\alpha}}-t = 0.
 \label{gauge}
 \end{equation}
 This gauge condition coincides with the classical solution of the Friedmann equation, giving the dependence of the scale factor $a$ on the cosmic time $t$.
 Then, on requiring the conservation of the gauge condition in time and using the equation
 \begin{equation}
 \frac{d\xi}{dt}=\frac{\partial \xi}{\partial t} + N\{\xi,H\} = 0,
 \label{lapse1}
 \end{equation}
 where the curly braces mean the Poisson brackets, one can easily see that the lapse function is equal to unity as it should be. 
If we solve the constraint (\ref{constr}) and use the gauge-fixing condition (\ref{gauge}) to define the time, the action (\ref{action-Ham}) can be written in the following form
\begin{equation}
 S = \int dt (p_{\phi}\dot{\phi}-H_{\rm phys}),
 \label{action1-Ham}
 \end{equation} 
where the physical Hamiltonian is
 \begin{equation}
 H_{\rm phys} = \frac{Mp_{\phi}}{3t}.
 \label{Hamilton2}
 \end{equation}
The corresponding Schr\"odiger equation is 
\begin{equation}
i\frac{\partial\psi_{\rm phys}(p_{\phi},t)}{\partial t} = H_{\rm phys}\psi_{\rm phys}(p_{\phi},t) = \frac{Mp_{\phi}}{3t}\psi_{\rm phys}(p_{\phi},t).
\label{Schrod100}
\end{equation}
The solution of this equation is 
\begin{equation}
\psi_{\rm phys}(p_{\phi},t) = \tilde{\psi}(p_{\phi})e^{-\frac{iMp_{\phi}}{3}\ln t}.
\label{Schrod102}
\end{equation}
Using Equation (\ref{gauge}), one can express the time $t$ as a function of the variables $\alpha$ and $p_{\phi}$; hence, we come back to one of the two branches of the general solution of the Wheeler-DeWitt equation (\ref{WdW-sol}). 
Let us note that the probability density $\tilde{\psi}^*\tilde{\psi}$ corresponding to the function (\ref{Schrod102}) does not depend on the cosmic time $t$, and this function can be normalized. 

Interestingly, we can take one of the branches of the general solution of the Wheeler-DeWitt equation and express the variable $\alpha$ as a function of time and of the variable 
$p_{\phi}$. On considering this function as the physical wave function satisfying the Schr\"odinger equation, we can calculate its partial time derivative to find 
the physical Hamiltonian. The results will coincide with those of Equation~(\ref{Hamilton2}). 
Note that, to obtain these results, we have used the super-Hamiltonian constraint.

\subsection{Introducing Time without Using the Super-Hamiltonian Constraint\label{3.3}}

{In Ref. \cite{Peres}, the author considers the same model in slightly different notations. We will analyze his approach using our notations. The main feature of the approach developed in Ref. \cite{Peres} is the requirement that the time parameter, as a combination of phase space variables, should be such that its Poisson bracket with the super-Hamiltonian is equal to $1$ not only on the constraint surface but on all of the phase space. Note that this is not the case for the time parameter chosen in Ref. \cite{we-IJMPD} and described in the preceding subsection. Let us now consider our arguments from this point of view and compare them with those in Ref. \cite{Peres}.} 

We would like to introduce a time parameter by identifying it with some function of phase space variables. It would be convenient to choose it in such a way 
that the time parameter coincided with the cosmic time, which is equivalent to the fixing of the lapse function $N = 1$.
{This fixation of the time parameter is equivalent to the gauge fixing of \mbox{the type}}
\begin{equation}
\chi(t,\alpha,p_{\alpha},\phi,p_{\phi}) = t-\tilde{\chi}(\alpha,p_{\alpha},\phi,p_{\phi})=0.
\label{gauge10}
\end{equation} 
By taking the time derivative of Equation~(\ref{gauge10}), we obtain the condition of the gauge conservation
\begin{equation}
1 = N\{\tilde{\chi},{\cal H}\},
\label{gauge1}
\end{equation}
where $\{,\}$ is the classical Poisson bracket.

Thus, to have $N =1$, we should choose the function $\tilde{\chi}$ in such a way that 
\begin{equation}
1 = \{\tilde{\chi},{\cal H}\}.
\label{gauge2}
\end{equation}
Obviously the choice of the function $\tilde{\chi}$ is not unique. We can choose a function that depends only on geometrical variables $\alpha$ and $p_{\alpha}$. Let us try the following:
\begin{equation}
 \tilde{\chi}(\alpha,p_{\alpha}) =- \frac{L^3M^2e^{3\alpha}}{3p_{\alpha}}.
 \label{gauge3}
 \end{equation}
As we have seen in Section~\ref{3.1}, the classical solution of the Friedmann equation of the model under consideration is 
\begin{equation}
a(t) = a_0t^{1/3},
\label{class}
\end{equation}
or 
\begin{equation}
\alpha(t) = \ln a_0+\frac13\ln t,\ \ \dot{\alpha} =\frac{1}{3t}.
\label{class1}
\end{equation}
Substituting Equation~\eqref{class1} into Equation~(\ref{momentum}) with $N=1$ gives 
\begin{equation}
p_{\alpha}=-\frac{L^3M^2a_0^3}{3}.
\label{class2}
\end{equation}
From \eqref{class1}, \eqref{class2}, and 
 \eqref{gauge3}, we can confirm that
\begin{equation}
 \tilde{\chi}(\alpha,p_{\alpha}) = t.
\label{time}
\end{equation}
Calculating the Poisson bracket of the function $\tilde{\chi}(\alpha,p_{\alpha})$ with the super-Hamiltonian (\ref{Hamilton10}), we obtain
\begin{equation}
\{\tilde{\chi},{\cal H}\} = \frac12\left(1+\frac{M^2p_{\phi}^2}{p_{\alpha}^2}\right),
\label{Poisson100}
\end{equation}
which is different from $1$. However, if we use the constraint ${\cal H} = 0$, we see that on the constraint surface our Poisson bracket is equal to $1$.
Now, we can consider the function $\tilde{\chi}(\alpha,p_{\alpha})$ as a new phase space variable $T$ and find its conjugate momentum. 
The momentum 
\begin{equation}
p_T = -\frac{p_{\alpha}^2e^{-3\alpha}}{L^3M^2}
\label{conjugate}
\end{equation}
satisfies the relation 
\begin{equation}
\{T,p_T\} = 1,
\label{Poisson10}
\end{equation}
as it should be. 
Besides, 
$$
\{T,\phi\}=\{T,p_{\phi}\} = \{p_T,\phi\} = \{p_T,p_{\phi}\} = 0.
$$
Now, we can make the canonical transformation from $\alpha,p_{\alpha}$ to $T,p_T$, without involving the variables $p_{\phi}$ and $\phi$.
This gives us an opportunity to write the reduced physical Hamiltonian as
\begin{equation}
H_{\rm phys} = -p_T = -\frac{p_{\alpha}^2e^{-3\alpha}}{L^3M^2}.
\label{Ham-phys}
\end{equation}
Using Equations (\ref{gauge3}), (\ref{time}), and (\ref{Hamilton10}), we can rewrite the expression (\ref{Ham-phys}) as follows (up to a sign)
\begin{equation} 
H_{\rm phys} = -p_T = \frac{Mp_{\phi}}{3t}.
\label{Ham-phys1}
\end{equation}
Thus, we have reproduced the formula obtained in Section~\ref{3.2}. 

In Ref. \cite{Peres}, Peres considered the same model (in slightly different notations and with different normalizations and variables).
He wanted to introduce the time parameter that would coincide with the cosmic time and looked for the function of the phase space variables whose Poisson bracket with the Hamiltonian 
is equal to $1$. This function would play the role of the cosmic time. To find this function, he solved the second-order equations of motion of the model, without using the super-Hamiltonian constraint (first-order equation of motion). Thus, his time parameter should be represented by the phase space variables everywhere and not only on the constraint surface. 
Otherwise, he says, the method would not be consistent. 

In practice, one can find the corresponding function without solving equations of motion, but by simply looking for the function of the phase space variables whose Poisson bracket with the super-Hamiltonian is equal to $1$ everywhere. Obviously, such a function cannot depend only on the geometrical phase variables $\alpha$ and $p_{\alpha}$, but should also depend on $p_{\phi}$. 
After some calculations, we find the following function
\begin{equation}
T_P = \frac{2M^2L^3e^{3\alpha}}{3(Mp_{\phi}-p_{\alpha})}.
\label{Peres}
\end{equation}
One can easily check that its Poisson bracket with the super-Hamiltonian is equal to unity in all of the phase space and not only on the constraint surface. 
If we resolve the constraint and choose the appropriate sign, i.e., $Mp_{\phi} = -p_{\alpha}$, \eqref{Peres} transforms into the function $T$ introduced earlier (see \eqref{gauge3}). 
There is also another function 
\begin{equation}
T_{P1} = -\frac{2M^2L^3e^{3\alpha}}{3(Mp_{\phi}+p_{\alpha})}.
\label{Peres1}
\end{equation}
 Its Poisson bracket with the super-Hamiltonian is identically equal to one. It coincides with our function $T$ if we resolve the constraint as $Mp_{\phi} = p_{\alpha}$. 
 
 Now, let us look for the momentum conjugate to the new phase space coordinate $T_P$: 
 by requiring that $\{T_P,P_{T_{P}}\} = 1$, we obtain
 \begin{equation}
 P_{T_{P}} = \frac{e^{-3\alpha}}{M^2L^3}\left(-\frac12p_{\alpha}^2+Ap_{\phi}^2+\left(\frac{M}{2}-\frac{A}{M}\right)p_{\phi}p_{\alpha}\right),
 \label{Peres2}
 \end{equation}
 where $A$ is a free real parameter. Note that, in Ref. \cite{Peres}, the existence of this additional freedom of choice is overlooked, and the conjugate momentum has a definite expression without free parameters.

In Ref. \cite{Peres}, the super-Hamiltonian is represented as a sum of two terms. One of these terms coincides with the momentum $P_{T_{P}}$ given by the expression (\ref{Peres2}), whereas the remaining term does not depend on $P_{T_{P}}$. It is this second term that plays the role of the physical Hamiltonian. Such an interpretation is based on the following reasoning. The action of the super-Hamiltonian on the quantum state is equal to zero. The conjugate momentum 
$P_{T_{P}}$ can be represented as 
\begin{equation}
P_{T_{P}} = -i \frac{\partial}{\partial T_{P}}.
\label{con-der}
\end{equation} 
Then,
from 
\begin{equation}
{\cal H}|\psi\rangle = (P_{T_{P}}+H_{\rm phys-P})|\psi\rangle=\left(-i \frac{\partial}{\partial T_{P}}+H_{\rm phys-P}\right)|\psi\rangle=0,
\label{con-der1}
\end{equation}
 it follows that 
 \begin{equation}
 i \frac{\partial}{\partial T_{P}}|\psi\rangle = H_{\rm phys-P}|\psi\rangle.
 \label{con-der2}
 \end{equation}
 Thus, we have obtained the Schr\"odinger equation where $T_P$ plays the role of time and $H_{\rm phys-P}$ is the physical Hamiltonian. Explicitly, 
 \begin{equation}
  H_{\rm phys-P}={\cal H} - P_{T_{P}} = \frac{e^{-3\alpha}(M^2-2A)p_{\phi}(Mp_{\phi}-p_{\alpha})}{2L^3M^3}.
  \label{Peres3}
  \end{equation}
  We can get rid of the factor $e^{-3\alpha}$ using our new variable $T_P$:
  \begin{equation}
  e^{-3\alpha} = \frac{2M^2L^3}{3T_P(Mp_{\phi}-p_{\alpha})}.
  \label{Peres4}
  \end{equation}
  Substituting the expression (\ref{Peres4}) into Equation~(\ref{Peres3}) gives the following expression for the physical Hamiltonian 
  \begin{equation}
  H_{\rm{phys-P}} = \frac{(M^2-2A)p_{\phi}}{3Mt},
  \label{Peres5}
  \end{equation}
where $t\equiv{T_P}$.
\\ 
The expression (\ref{Peres5}) has some interesting features. 
\begin{enumerate}
\item 
It has the same structure as the expression (\ref{Ham-phys1}); namely, it is proportional 
to the momentum $p_{\phi}$ and inversely proportional to the cosmic time parameter $t$. 
\item 
The expression for the time parameter as a function of phase space variables was obtained without using the constraint ${\cal H} = 0$. 
\item 
The combination of the phase variables representing the cosmic time parameter was chosen in such a way that its Poisson bracket with the super-Hamiltonian is equal to $1$ in all of the phase space and not only on the constraint surface. 
\item 
There is a free parameter $A$ in the expressions (\ref{Peres2}) and (\ref{Peres5}). The existence of such a freedom was not 
noticed in Ref. \cite{Peres}. Indeed, one can see that the expression for $P_{T_{P}}$ and, hence, for $H_{\rm phys-P}$ in Ref. \cite{Peres}, corresponds to
 the choice 
$$
A = -\frac{M^2}{2}.
$$
Our physical Hamiltonian (\ref{Ham-phys1}) arises if we choose
$$
A = 0.
$$
\item 
Although we did not use the classical constraint to introduce the time parameter as a function of the phase space variables, when defining the physical Hamiltonian, we~used the fact that the quantized super-Hamiltonian eliminates the physical state. 
\end{enumerate} 

Let us stress that it is our physical Hamiltonian \eqref{Ham-phys1} that is compatible with the corresponding classical evolution of the model. 

So far, we have demonstrated that the whole family of reduced physical Hamiltonians, which depend on a free parameter $A$, corresponds to the same choice of the classical cosmic time parameter. How can this parameter be fixed?
 Let us write the Schr\"odinger equation for the Hamiltonian (\ref{Peres5}):
\begin{equation}
i\frac{\partial\psi(p_{\phi},t)}{\partial t} = \frac{(M^2-2A)p_{\phi}}{3Mt}\psi(p_{\phi},t).
\label{Schrod101}
\end{equation} 
Its solution is 
\begin{equation}
\psi(p_{\phi},t) = \psi(p_{\phi}) \exp\left(-i\frac{(M^2-2A)p_{\phi}}{3M}\ln t\right),
\label{Schrod2}
\end{equation}
where the function $\psi(p_{\phi})$ should satisfy the normalizability condition
\begin{equation}
\int dp_{\phi}\psi^*(p_{\phi})\psi(p_{\phi}) = 1.
\label{Schrod3}
\end{equation}
The expectation value of the momentum
\begin{equation}
\langle p_{\phi} \rangle \equiv \langle \psi(p_{\phi},t)|p_{\phi}|\psi(p_{\phi},t)\rangle = \int dp_{\phi}p_{\phi}\psi^*(p_{\phi})\psi(p_{\phi}) 
\label{Schrod4}
\end{equation}
does not depend on time, just like its classical analogue. What happens with the expectation value of the scalar field $\phi$?
\begin{eqnarray}
\langle \phi \rangle \equiv \langle \psi(p_{\phi},t)|\phi|\psi(p_{\phi},t)\rangle&=&i\int dp_{\phi}\psi^*(p_{\phi},t)\frac{\partial\psi(p_{\phi},t)}{\partial p_{\phi}}\nonumber\\&=&\langle \phi \rangle_{t=1}+\frac{M^2-2A}{3M} \ln t,
\label{Schrod5}
\end{eqnarray}
where $$\langle \phi \rangle_{t=1}\equiv i\int dp_{\phi}\psi^*(p_{\phi},t)\frac{d\psi(p_{\phi},t)}{d p_{\phi}}$$ is the expectation value at $t=1$.

Let us now find the classical evolution of the scalar field $\phi$. 
Using the Lagrangian 
(\ref{Lagrange1}) and making variation with respect to the lapse function $N$, we obtain the following constraint in the Lagrangian formalism
\begin{equation}
M^2\dot{\alpha}^2 = \dot{\phi}^2.
\label{class-constr}
\end{equation} 
Note that this constraint does not depend on the choice of the time parametrization. 
Let us choose the solution as follows
\begin{equation}
\dot{\phi} = M\dot{\alpha}.
\label{class-constr1}
\end{equation}
We have already mentioned that $\dot{\alpha} = \frac{1}{3t}$ (see Equation~(\ref{class1}). Thus,
\begin{equation}
\dot{\phi} = \frac{M}{3t}.
\label{class-constr2}
\end{equation} 
 We see that (\ref{class-constr2}) and (\ref{Schrod5}) are compatible if $A=0$, i.e., if the 
reduced Hamiltonian is equal to our Hamiltonian (\ref{Ham-phys1}).

\subsection{Hamilton-Jacobi Type Equation and the Choice of the Time Parameter and of the Hamiltonian\label{3.4}}

As a matter of fact, the author of Ref. \cite{Peres} gives his definition of time and of the Hamiltonian using the solution of the Hamiltonian-Jacobi type equation and not the method of direct construction of new canonical variables, which we used in the preceding subsection. One can try to find out if the solution of the corresponding partial differential equation can fix these variables in an unambiguous way. We will show in this subsection that this not the case and that the arbitrariness, which was not noticed in Ref. \cite{Peres}, is still present.

Let us introduce the volume variable
\begin{equation}
v \equiv a^3 \equiv e^{3\alpha}.
\label{volume}
\end{equation}
In terms of this variable, the Lagrangian \eqref{Lagrange} is as follows:
\begin{equation}
{\cal L}  =  -\frac{L^3M^2\dot{v}^2}{18Nv}+\frac{L^3\dot{\phi}^2 v}{2N}. 
\label{volume1}
\end{equation}
We can put $N=1$ because we will be looking for the time parameter that coincides with the cosmic time (the author of Ref. \cite{Peres} calls it ``auxiliary time $\tau$''). 
After this, we can write the Euler-Lagrange equation with respect to $v$, which is nothing but the second Friedmann equation:
\begin{equation}
2\ddot{v}v-\dot{v}^2+\frac{9p_{\phi}^2}{L^3M^2}=0,
\label{volume2}
\end{equation}
where we have used the relation
\begin{equation}
\dot{\phi}=\frac{p_{\phi}}{L^3v}.
\label{volume3}
\end{equation}
Now, we can find the solution of the second Friedmann Equation \eqref{volume2} without using the first Friedmann equation (the super-Hamiltonian constraint). With the initial condition $v(0) = 0$, the solution is 
\begin{equation}
v(t) = Kt^2+3\frac{p_{\phi}}{L^3M}t,
\label{volume4}
\end{equation}
where \emph{K} is an arbitrary constant. 
Using the expression \eqref{volume4}, we can calculate explicitly the expression for the super-Hamiltonian.
Indeed, 
\begin{equation}
\dot{v} = 2Kt+3\frac{p_{\phi}}{L^3M}.
\label{volume5}
\end{equation}
Then, 
\begin{equation}
p_v = -\frac{L^3M^2}{9}\frac{\dot{v}}{v},
\label{volume6}
\end{equation}
and 
\begin{equation}
H=-\frac{9}{2M^2L^3}vp_v^2+\frac{1}{2L^3}\frac{p_{\phi}^2}{v}=-\frac29KL^3M^2.
\label{volume7}
\end{equation}
Thus, we see that the super-Hamiltonian does not depend on time and is proportional to the constant $K$.
If we take into account the first Friedmann equation, or, in other words the super-Hamiltonian constraint, the constant 
$K$ should be equal to zero. Note that the dependence of the volume on time given by Equation \eqref{volume4}
arises naturally in the universe filled with dust and a massless scalar field, which behaves as a stiff matter (see, e.g., Ref. \cite{Kam-Khal}). On the other hand, this behavior arises in the generalized unimodular theory of gravity~\cite{gen-unimod},
where the lapse function is treated not as a Lagrange multiplier but as a certain function of the determinant of the spatial metric. If one chooses the lapse function to be a constant (or in other words, chooses a gauge fixing condition corresponding to the cosmic time), an effective dust arises in the universe. If the lapse function is inversely proportional to the determinant of the spatial metric, one has the unimodular theory of gravity, and an effective cosmological constant appears~\cite{unimod,unimod1}.

Now, using Equations \eqref{volume4}--\eqref{volume6}, we can easily express the time parameter $t$ as a function of the phase variables $v, p_{v}$ and $p_{\phi}$:
\begin{equation}
t = \frac{2M^2L^3v}{3(Mp_{\phi}-3vp_v)}.
\label{volume8}
\end{equation} 
It is easy to check that this expression coincides with the Formula \eqref{Peres}, which we obtained by simply looking for the combination of the phase variables whose Poisson bracket with the super-Hamiltonian equals to $1$ in all of the phase space. Note that the constant $K$ does not enter into this expression.

The next step is to find the canonical momentum conjugate to the new coordinate given by the expression \eqref{volume8}. In the preceding subsection, we did it using the definition of the Poisson bracket. In Ref. \cite{Peres}, the author constructed the corresponding canonical transformation. For this purpose the generating function of the type $F_1$ (see, e.g., Ref. \cite{Goldstein}) was used. The generating function can be written as $S(q,Q)$, where $q$ are the old coordinates, while $Q$ are the new coordinates. Then, 
\begin{equation}
p_{k} = \frac{\partial S}{\partial q^k},
\label{canon}
\end{equation}
\begin{equation}
P_{\mu} = \frac{\partial S}{\partial Q^{\mu}}.
\label{canon1}
\end{equation}

Now, as a first new coordinate we choose the coordinate $T$, which coincides with the expression of the right-hand side of Equation \eqref{volume8}. It will be convenient for us to go back to our old variable $\alpha$. Thus,
\begin{equation}
T = \frac{2M^2L^3e^{3\alpha}}{3(Mp_{\phi}-p_{\alpha})}.
\label{canon2}
\end{equation}
As a second new coordinate we choose 
\begin{equation}
Q = p_{\phi}.
\label{canon3}
\end{equation}
Obviously, the Poisson bracket between the expressions \eqref{canon2} and \eqref{canon3} is equal to zero.
From Equation \eqref{canon2}, we have 
\begin{equation}
p_{\alpha} = MQ-\frac23\frac{M^2L^3e^{3\alpha}}{T}.
\label{canon4}
\end{equation}
The generating function for which we are looking depends on four variables:
\begin{equation}
S = S(\alpha,\phi, T,Q).
\label{canon5}
\end{equation}
Thus,
\begin{equation}
p_{\alpha} = \frac{\partial S}{\partial \alpha}=MQ-\frac23\frac{M^2L^3e^{3\alpha}}{T}.
\label{canon6}
\end{equation}
The general solution of this equation is 
\begin{equation}
S(\alpha,\phi,T,Q) = MQ\alpha-\frac29\frac{M^2L^3e^{3\alpha}}{T}+\tilde{S}(\phi,T,Q).
\label{canon7}
\end{equation}
We can impose an additional constraint on the function $\tilde{S}$. The partial derivative of the generating function with respect to $\phi$ should give us $p_{\phi}=Q$. Hence,
\begin{equation}
\frac{\partial S}{\partial \phi} = \frac{\partial \tilde{S}}{\partial \phi} = p_{\phi} = Q.
\label{canon8}
\end{equation}
Then,
\begin{equation}
\tilde{S} = \phi Q+\bar{S}(T,Q).
\label{canon9}
\end{equation}
We still have some freedom of choice for the function $\bar{S}(T,Q)$. To make the comparison with the results of the preceding two subsections simpler, let us choose it as follows:
\begin{equation}
\bar{S}(T,Q) = \frac23\left(\frac{M}{2}+\frac{A}{M}\right)Q\ln T,
\label{canon10}
\end{equation}
where $A$ is an arbitrary constant. 

Now, we can find the momentum $P_T$ using the Formula \eqref{canon1}:
\begin{equation}
P_T = -\frac{\partial S}{\partial T} = -\frac29\frac{M^2L^3e^{3\alpha}}{T^2}-\frac23\left(\frac{M}{2}+\frac{A}{M}\right)\frac{Q}{T}.
\label{canon11}
\end{equation} 
It is easy to check that, if we go back to the old phase space variables $\alpha,\phi, p_{\alpha}, p_{\phi}$, the formula \eqref{canon11}
coincides with the Formula \eqref{Peres2}. Thus, using the canonical transformation and the generating function we came to the same result as was obtained by the direct study of the corresponding Poisson brackets. Naturally, the effective physical Hamiltonian is given by the Formula \eqref{Peres5}.

\subsection{Can the Choice $A \neq 0$ Correspond to Some Classical Evolution?\label{3.5}}

At the end of Section~\ref{3.3}, we saw that, if $A$ is different from zero (for example, if $A=-\frac{M^2}{2}$, as was implicitly chosen in Ref. \cite{Peres}), then the quantum evolution is not compatible with the classical: the time derivative of the expectation value of the field is given by 
\begin{equation} 
\frac{d\langle\phi\rangle}{dt}=\frac{M^2-2A}{3Mt},
\label{quant-class}
\end{equation} 
instead of 
\begin{equation}
\frac{d\langle\phi\rangle}{dt}=\frac{M}{3t}.
\label{quant-class1}
\end{equation}
We can ask if the classical counterpart of \eqref{quant-class} exists in the theory where the super-Hamiltonian constraint is not imposed. To answer this question, let us recall that the relation between the momentum $p_{\phi}$, the time derivative of the scalar field $\dot{\phi}$ and the volume $v = a^3=e^{3\alpha}$, 
\begin{equation}
p_{\phi} = L^3v\dot{\phi},
\label{quant-class2}
\end{equation} 
does not depend on imposing the super-Hamiltonian constraint and follows simply from the Klein-Gordon equation for the scalar field. By combining Equations \eqref{quant-class} and \eqref{quant-class2}, we obtain
\begin{equation}
v(t) = \frac{3p_{\phi}M}{L^3(M^2-2A)}t.
\label{quant-class3}
\end{equation}
On the other hand, from the second Friedmann equation and the Klein-Gordon equation, we obtain the expression \eqref{volume4}. We see that it cannot coincide with \eqref{quant-class3} unless $A = 0$.

\section{Switching Internal Times and the Montevideo Interpretation of Quantum Theory}

In the recent paper by Ref. \cite{Hohn}, the author considers the same simple model that we discussed in the preceding section. His approach to the problem of time and related problems in quantum gravity was laid out in Ref. \cite{Hohn1, Hohn2, Hohn3, Hohn4}. The author states that the question of the quantum notion of general-relativistic covariance received little attention; he addresses this question by using quantum reference  frames. One of the main points of Ref. \cite{Hohn} is the explicit and detailed study of switching between different internal times. The toy model is very convenient from this point of view. Indeed, all the calculations can be done analytically, and the obtained formulae are rather simple. There is also another advantage: the symmetry between the logarithm of the scale factor and the scalar field. This symmetry is especially clear if one looks at the super-Hamiltonian constraint \eqref{Hamilton10}--\eqref{constr}, which we can 
conveniently rewrite as follows:
 \begin{equation}
 p_{\alpha}^2-M^2p_{\phi}^2=0.
 \label{swit}
 \end{equation}

Generally, we find an essential correspondence between the approaches of Ref. \cite{Hohn,Barv,Bar-Kam,we-IJMPD} and Section~\ref{3.2} of the present paper. Indeed,~Ref.\cite{Hohn} uses a method that ``identifies a consistent reduction procedure that maps the Dirac quantized theory to the various reduced quantized versions of it to different choices of quantum reference  systems. It identifies the physical Hilbert space of the Dirac quantization as a reference-system-neutral-quantum superstructure and the various reduced quantum theories as the physics described relative to the corresponding choice of reference  system.'' Hence, the quantum state annihilated by the Dirac constraint (or, in other words, satisfying the Wheeler-DeWitt equation) does not have a direct physical interpretation. This opinion coincides with ours. We also stressed that the physical (probabilistic) interpretation arises when we construct the time parameter from a part of the phase space variables and this parameter, as well as its conjugate momentum, are excluded from the effective Hamiltonian by using the constraint.
Let us note that the author of Ref. \cite{Hohn} also uses the constraint at all the stages of his procedure, in contrast to the author of Ref. \cite{Peres}.

Because of the symmetry of our model, there are four natural ways to introduce a time parameter or, in other words, a quantum clock~\cite{Hohn}. One can choose a function of the scale factor as a time variable, and, in this case, the wave function will depend on the scalar field, or one can consider the scalar field as a quantum clock. Two other choices arise when we use Eq. \eqref{swit} to exclude the momentum conjugate to the time variable from the effective Hamiltonian. In the language of Ref. \cite{Hohn}, it is called ``trivialize the constraint to the internal time to render it redundant''.

In Section~\ref{3.2}, the function of the scale factor was chosen to be the internal time. 
Let us try to describe the switching between different internal times in our terms.
We assign the role of time to a suitable function of the scalar field $\phi$. Solving jointly the Friedmann equation \eqref{Fried1} and the Klein-Gordon equation \eqref{KG} gives
\begin{equation}
\phi(t) = \frac{M}{3}\ln\frac{t}{t_0},
\label{swit1}
\end{equation} 
where $t_0$ is a constant. Now, we can introduce a new canonical variable
\begin{equation}
T=T_0\exp\left(\frac{3\phi}{M}\right);
\label{swit2}
\end{equation}
its conjugate momentum is
\begin{equation}
p_T= \frac{Mp_{\phi}}{3T_0}\exp\left(-\frac{3\phi}{M}\right)=\frac{Mp_{\phi}}{3T}.
\label{swit3}
\end{equation}
 Using the algorithm described in Section~\ref{3.2} and the constraint \eqref{swit}, we obtain the following effective physical 
Hamiltonian:
\begin{equation}
H_{{\rm phys}-p_{\alpha}} = \pm \frac{p_{\alpha}}{3t},
\label{swit4}
\end{equation}
where the sign ``$\pm$'' corresponds to the two options existing in the quadratic form of the constraint \eqref{swit}. 
It is easy to see that the form of the Hamiltonian \eqref{swit4} is quite similar to that of \eqref{Hamilton2}. Likewise, the solution of the corresponding Schr\"odinger equation $\psi(p_{\alpha},t)$ has the structure quite similar to that of \eqref{Schrod102}.
The switching between two choices of time gives similar results due to the symmetry and simplicity of the model as was noted in Ref. \cite{Hohn}. The two options---the time related to purely geometric characteristic of the universe and the time connected with the scalar field---look quite natural in this model. However, other choices also exist. For example, instead of $\alpha$ and $\phi$, we can introduce a different pair of canonical variables:
\begin{eqnarray}
&&T = T_0\exp\left(\frac{3}{2M}(\phi+M\alpha)\right),\nonumber \\
&&\beta = M\alpha.
\label{swit5}
\end{eqnarray}
The conjugate momenta are
\begin{eqnarray}
&&p_T=\frac{2Mp_{\phi}}{3T_0}\exp\left(-\frac{3}{2M}(\phi+M\alpha)\right),\nonumber \\
&&p_{\beta}=-p_{\phi}+\frac{p_{\alpha}}{M}.
\label{swit6}
\end{eqnarray}
Implementing the procedure described above, we arrive at the effective Hamiltonian
\begin{equation}
H_{{\rm phys}-p_{\beta}} = \frac{Mp_{\beta}}{3t},
\label{swit7}
\end{equation}
which again has the same form as Hamiltonians considered before.

We should mention that the internal time parameters that do not mix the geometrical variable $\alpha$ and the matter variable $\phi$ look more natural and are more convenient for work with. However, it is not always so. When one considerers the models where the scalar field is non-minimally coupled to gravity, it is more convenient to introduce an internal time parameter that mixes the geometrical and the matter variables; see, e.g., Ref. \cite{KTV,KTV1}.

Further elaboration of the procedure of switching between internal times~\cite{Hohn5,Hohn6} has some implications for the Montevideo interpretation of quantum mechanics~\cite{Mont}. To discuss this topic, let us first recapitulate the main features of this interpretation~\cite{Mont0,Mont1,Mont2,Mont3,Mont4}.

As with all interpretations of quantum mechanics, the main target of the Montevideo interpretation is the problem of measurement. This problem can be formulated as the necessity to understand how the principle of superposition and the linearity of the Schr\"odinger equation can be reconciled with the fact that, in every measurement, we see only one outcome. The Copenhagen interpretation suggests that there are two realms---the quantum and the classical---and there is a complementarity between them. The Montevideo interpretation is based on the principle that the quantum description of reality is the only fundamental one. This is akin to the Many-Worlds interpretation of quantum mechanics~\cite{Everett,many-worlds}. The distinguishing feature of the Montevideo interpretation is its attention to the notion of time in quantum theory. The clear distinction is made between the quantum clock, which is connected with some quantum variable and is an operator, and the coordinate time. It is important that there are quantum limits on the precision of the time measurement by quantum clocks. First of all, there is the well-known time-energy uncertainty relation or Mandelstam-Tamm relation~\cite{Tamm}. Besides, the presence of gravity implies that there are additional constraints on the resolution of quantum clocks. Thus, according to the Montevideo interpretation, while the quantum system has a unitary evolution with respect to the coordinate time, its evolution with respect to the quantum clock is non-unitary, and a pure state transforms into a mixed one. This means that the corresponding density matrix evolves not according to the quantum Liouville (von Neumann) equation but according to the equation with some additional terms of the Lindblad type~\cite{Lindblad,Gorini}. This equation has the following form
\begin{equation}
\frac{\partial\rho(T)}{\partial T} = i[\rho(T),H]+\sigma(T)[H,[H,\rho(T)]]+\cdots,
\label{Mont}
\end{equation}
where $T$ is a quantum clock time. It is written in Ref. \cite{Mont} that the Inclusion of the of quantum gravitational notion of time in the spirit of the papers~\cite{Hohn5,Hohn6} makes the construction based on the Montevideo interpretation more solid. In this connection we see an interesting technical problem. Can we starting with different definitions of internal time, like those considered in Ref. \cite{Hohn} and in the present paper, to obtain the forms of the function $\sigma(T)$ responsible for the non-unitary evolution of the density matrix? It would be curios to see how it works at least for simple toy models.

Before finishing this section, we would like to say once again that from the point of view of classification of different interpretations of quantum mechanics, the Montevideo interpretation is rather close to the Everett interpretation but with a special emphasis on the role of time and gravitational effects. Besides, the branching that arises in the Everett interpretation as a result of a measurement-like process becomes approximate. On the other hand, the Montevideo interpretation can be considered as one of the approaches to the synthesis of gravity and quantum theory. Investigations of a possible influence of gravitational effects on quantum physics have a rather long history~\cite{Wigner,Karolyhazy,Diosi,Penrose,Kibble,Kibble1}. The authors proposing the Montevideo interpretation stress that they do not modify quantum mechanics. They simply consistently take into account the gravitational effects, which results in resolution of some long-standing problems of quantum theory. An interesting example of quite a different attitude to the problem of synthesis between quantum theory and gravity is the recently proposed Correlated Worldline Theory~\cite{Stamp1,Stamp2,Stamp3,Stamp4}. In the framework of this theory, the generalized equivalence principle is implemented and new path integrals representing the generating functionals are obtained. The linearity of the quantum theory disappears, but is restored in the limit when the gravitational interaction is switched off. Thus, we see another interesting problem here: what happens with the notion of time in this new theory?


\section{Discussion}

In this review various ideas concerning the problem of time in quantum gravity and cosmology were studied by using a simple model of a flat Friedmann universe filled with a massless minimally coupled scalar field. This model was studied in a relatively old paper, Ref. \cite{Peres}, and in some new works~\cite{Bar-Kam,we-IJMPD,Hohn}. Why does this model attract the attention of researchers? It is very simple, and all the calculations can be done analytically for classical and quantum dynamics. At the same time, most of the problems can be seen by looking at this simple example. What would change if we included a non-vanishing mass? As we can see from the pioneering paper of Ref. \cite{BGKZ}, even classical dynamics becomes very rich and there are no exact solutions. One can also consider scalar fields with other potentials. For example, for a constant potential one can find the exact solutions of the Wheeler-DeWitt equation~\cite{Bar-Kam}; the same is true for an exponential potential~\cite{Andrianov}. However, these more complicated minisuperspace models give almost no additional insight. The same can be said about more realistic models, which contain anisotropies or/and perturbations. To introduce an internal time or a quantum clock, one usually uses macroscopic variables, such as the scale factor and the homogenous mode of a scalar field. Thus, one can try to understand the basic features of the internal time by playing with relatively simple models, which we tried to do in the present review.

Upon analyzing the method of introduction of time without using the super-Hamiltonian constraint, we found a flaw in the calculations of Ref. \cite{Peres} and demonstrated that some freedom of choice of variables was overlooked. One can represent (at least partially, but it was enough for our analysis) this freedom by a single parameter $A$. Choosing this parameter as $A=0$ gives the evolution of the quantum average of the scalar field that coincides with its classical counterpart; with any other choice (including the one implicitly made in the Ref. \cite{Peres}) there is not a sensible classical counterpart. On the other hand, the correct expression for the effective Hamiltonian was obtained in the framework of the formalism where the super-Hamiltonian constraint was used at all the stages of the problem~\cite{Barv,Bar-Kam,we-IJMPD}. Thus, it can be concluded that the introduction of new variables with the desired value of the Poisson brackets in all of the phase 
 space is neither necessary nor sufficient for a reasonable treatment of the Wheeler-DeWitt equation. 

Nevertheless, the paper written by A. Peres is, certainly, thought-provoking because it makes the reader think about such complicated questions as the problem of time in quantum cosmology, the correct treatment of the Wheeler-DeWitt equation, the transition to the Hamiltonian formalism in theories that possess reparametrization invariance, and, last but not least, the complicated interrelations between the classical and quantum theories. The author of Ref. \cite{Peres} advocates the Copenhagen interpretation of quantum mechanics. He writes: ``On the other hand, quantum theory, unlike general relativity, is not a ``theory of everything''. Its mathematical formalism can be given a consistent physical
interpretation only by arbitrarily dividing the physical world into two parts: the
system under study, represented by vectors and operators in a Hilbert space, and
the observer (and the rest of the world), for which a classical description is used''. The Ref.  to the famous paper by Niels Bohr~\cite{Bohr} is given in the Introduction of Ref. \cite{Peres}.

In all the calculations and considerations of Section~\ref{3} of the present paper, the treatment of time as a classical variable and the comparison between the classical and quantum evolutions are used. At the same time, a different point of view becomes more and more popular among researchers working in quantum cosmology. It is related to the many-worlds interpretation of quantum mechanics~\cite{Everett, many-worlds} and to the idea that the quantum theory is the only fundamental theory. The classical behavior emerges only in some limits and contexts (see, e.g.,~Ref. \cite{Bar-Kam-Zeh} for a recent review). If we accept the fundamental character of the quantum theory, we should expect the appearance of some quantum properties of time, considering that the time parameter in quantum gravity and cosmology is a combination of quantum operators. Here, this ``quantumness'' was in a way frozen, but this phenomenon can hardly have a universal character. Is it possible to see 
 some flashes revealing the true quantum nature of time?

Let us note that ideas concerning this quantum nature of time were already discussed in some works. For example, in Ref. \cite{TVV-PRD}, a simple cosmological model, similar to that in the present paper but with the massive scalar field, was considered. Using the Born-Oppenheimer approach, the authors factorized the solution of the Wheeler-DeWitt equation into two parts: one depending only on the geometrical variables (scale factor), and the other depending also on the matter variables and the perturbations of the homogeneous background. The purely geometrical part of the wave function is responsible for the emergence of time, while the other part describes the quantum evolution of the matter degrees of freedom with respect to this time parameter. It was discovered that, to make this definition of time reasonable, it is necessary to use some kind of course-graining procedure. Without this procedure the time variable behaves in a strange, oscillatory way.

The idea of Ref. \cite{TVV-PRD} was further developed in Ref. \cite{we-PRD}, where the attempt to construct a ``non-semiclassical'' wave function of the universe was undertaken, and the possible observable effects, such as the influence of this non-classicality on the spectrum of the cosmic microwave background, were predicted.

Some interesting features of time and causality in quantum gravity were studied in Ref. \cite{Anselmi,Anselmi1,Anselmi2}. As is known, when introducing the higher-derivative terms in the gravitational action to make it renormalizable, one encounters the problem of unitarity. To resolve this problem, the author of Ref. \cite{Anselmi,Anselmi1,Anselmi2} introduduces particles which he calls ``fakeons''. The propagators of these particles are not of Feynman type, and the standard notion of causality is not valid for them. Thus, one can say they live in a ``different time''. This is even more evident in Ref. \cite{Don-Men,Don-Men1}. To study the structure of the propagator of the graviton, the authors introduce the notion of the so-called ``Merlin modes'', which live in a time that runs backwards. These states are highly unstable, their appearance does not break the causality on the observable distances.

In our opinion, the question of genuine quantum nature of time and its relation to the classical evolution is one of the most interesting and challenging problems of quantum~cosmology.

\vspace{6pt}

All authors contributed equally to the present work. All authors have read and agreed to the published version of the manuscript.

This research was supported by the Russian Foundation for Basic Research grant No 20-02-00411 and 
by the research grant ``The Dark universe: A Synergic Multimessenger Approach'', No. 2017X7X85K under the program PRIN 2017 funded by the Ministero dell'Istruzione, delle Universit\`a e della Ricerca (MIUR), Italy.

We are grateful to A.O. Barvinsky, A. Tronconi and G. Venturi for fruitful discussions.

The authors declare no conflict of interest.

 
\end{document}